\documentstyle[12pt]{article}
\begin{document}

 \newcommand{\ID}{1 \! \! 1}
\newcommand{\fracts}[2]{\textstyle \frac{#1}{#2} }
\newcommand{\fracds}[2]{\displaystyle \frac{#1}{#2} }
%
%
 \renewcommand{\theequation}{\arabic{section}.\arabic{equation}}
%
 \newcommand{\sectione}[1]{ \section{#1}
                            \setcounter{equation}{0} }
%
 \newcommand{\startapp}{ \appendix
 \renewcommand{\theequation}{\Alph{section}.\arabic{equation}} }
\def\be{\begin{equation}}
\def\ee{\end{equation}}
\def\l{\lambda}
\def\L{\Lambda}
\def\pa{\partial}
\def\dd{\delta}
\def\de{\Delta}
\newcommand{\bea}{\begin{equation} \begin{array}{rcl}}
\newcommand{\eea}{\end{array} \end{equation}}
\newcommand{\dio}[1]{{\rm Diff}(#1)}

 \begin{titlepage}
 \vspace*{-4ex}
 \null \hfill MPI-Ph/92-25 \\
 \null \hfill UCB-PTH-92/10 \\
 \null \hfill LBL-32217  \\
%
 \begin{center}
 {\bf \LARGE Reality in the Differential Calculus on
}\\ [3ex]
{\bf \LARGE q-euclidean Spaces} \\  [8ex]
  O.\, Ogievetsky$^\dagger$
 \footnote{On leave of absence from P.N.Lebedev Physical
 Institute, Theoretical Department, 117924 Moscow, Leninsky
 prospect 53, CIS},
 and B.\, Zumino$^\ddagger$
\footnote{This work was supported in part by the Director,
 Office of Energy Research,
 Office of High Energy and Nuclear Physics,
 Division of High Energy Physics of the U.S. Department of Energy
 under Contract DE-AC03-76SF00098
 and in part by the National Science
 Foundation under grant PHY-85-15857}
   \\ [4ex]

 $^\dagger$ Max-Planck-Institut f\"{u}r
 Physik und Astrophysik \\
  - Werner-Heisenberg-Institut f\"{u}r Physik \\
 P.O.\,Box 40 12 12 , D - 8000 Munich 40, Germany \\ [2ex]
 $^\ddagger$ Department of Physics \\
    University of California, Berkeley \\
     and \\
    Theoretical Physics Group \\
    Physics Division \\
   Lawrence Berkeley Laboratory \\
  1 Cyclotron Road \\
  Berkeley, California 94720 \\ [8ex]
 \end{center}
\bigskip
 \begin{abstract}
The nonlinear reality structure of the
derivatives and the differentials
for the euclidean q-spaces are found. A real
Laplacian is constructed and reality properties of the
exterior derivative are given.
 \end{abstract}
 \end{titlepage}

\section{Introduction}
\vskip 0.3cm

In this paper we discuss a new
effect appearing in the differential
calculus on euclidean q-spaces. Namely, although the
conjugation rules for the coordinates look like those in
the classical case, the conjugation of the derivatives and the
differentials turns out to be nonlinear.
For the Minkowski q-space these conjugation rules were
discussed in \cite{OSWZ}. Here we generalize the results of
\cite{OSWZ} to the higher dimensional euclidean
q-spaces. The nonlinearity in the relation between the
derivatives and their
conjugates turn
out to be quite simple. Classically, the derivatives
are proportional to the commutator of the Laplacian with
the coordinates. On the quantum level these are however
two different objects.
Our main result is that the commutator of
the Laplacian with the coordinates is now proportional
to the conjugates of the derivatives. The coefficient of
the proportionality is no longer a number. It is a scaling
operator introduced in \cite{OG}. It q-commutes
with all the coordinates and derivatives.

Our treatment relies on the papers \cite{frt},
\cite{WZCOV}. The prescription for the differential calculus
on q-spaces, given in \cite{WZCOV} works when the
commutation relations for coordinates of a q-space
are given by a single projector entering the $\hat{R}$-matrix.
This is the case for q-orthogonal spaces.
Although the $\hat{R}$-matrices
have different structure for even and
odd dimensional q-orthogonal spaces,
the results have the same form
in both cases and we treat them simultaneously.

The paper is organized as follows.
Section 2 contains basic facts about the
orthogonal q-spaces and differential calculus on them.
In section 3 we give the
reality structure for the derivatives and discuss reality
properties of the Laplacian. Section 4 is devoted to the
differentials and reality properties of the exterior derivative.
Also, there we comment briefly on the relation between two
versions of the differential calculus in the SL$_q$ case.
In Appendix A we collected relevant relations used in the text.

The q-metric
$g_{ij}$
is not symmetric and throughout the text we
use the following rules of lowering and raising an index of
any one-index quantity $m_i$:
\be
m^i=g^{ij}m_j,\ m_i=g_{ij}m^j.
\ee
Conjugation reverses the order of factors.
Finally, we use the notation $\l = q-q^{-1}$.

\sectione{Preliminaries}
\vskip 0.3cm

Here we list necessary facts about the $\hat{R}$-matrix for
the orthogonal q-group $SO_q(N)$, euclidean q-spaces
and differential calculus on q-spaces. For motivations and details
we refer to \cite{frt},\cite{WZCOV}. See also \cite{sl} for the
discussion of the differential calculus on orthogonal
q-spaces.

1. The projector decomposition of the $\hat{R}$-matrix for the
orthogonal q-group $SO_q(N)$ is:
\be
\hat{R}=qP^+ -q^{-1}P^- + q^{1-N}P^o.
\label{ap1}
\ee
Here $P^+$ is the traceless part of the q-analogue of the
symmetriser, $P^-$ is the q-analogue of the
antisymmetriser and $P^o$ is the
trace projector. The projector $P^o$ is built out of the
q-metric $g_{ij}$,
\be
P^{o\ ij}_{\ \ kl}=\nu g^{ij}g_{kl}\ ,\ \ \ \ \ \nu =\frac{\l}{(
q^N-1)(q^{1-N}+q^{-1})} \ .
\label{ap2}
\ee
The $\hat{R}$-matrix has the following symmetry properties:
\be
\hat{R}^{-1\ ij}_{\ \ \ \ kl}=g^{im}\hat{R}^{jn}_{mk}g_{nl}=
g_{km}\hat{R}^{mi}_{ln}g^{nj}
\label{ap3}
\ee
and
\be
\hat{R}^{ij}_{kl}=\hat{R}^{kl}_{ij}.
\label{ap4}
\ee

2. The orthogonal q-space is the algebra with generators
$x^i,\ i=1,...,N$ satisfying quadratic relations
\be
 P^{- \ ij}_{\ \ \ kl} x^k x^l = 0
\label{ap5}
\ee
or
\be
\hat{R}^{ij}_{kl}x^k x^l =qx^i x^j -
\fracds{\l}{1+q^{N-2}} g^{ij}g_{kl}x^k x^l \ .
\ee
The length
\be
L=\frac{1}{1+q^{N-2}} g_{ij}x^i x^j
\label{ap6}
\ee
is the central element in the algebra of the coordinates,
$Lx^i = x^i L$.

The projectors $P^+ ,P^o$ define the quadratic relations for the
differentials $\xi^i$:
\be
P^{+\ ij}_{\ \ \ kl}\xi^k \xi^l =0,
\label{ap7}
\ee
\be
P^{o\ ij}_{\ \ kl}\xi^k \xi^l =0.
\label{ap8}
\ee
The derivatives $\pa_i$ are defined by
\be
\pa_i x^j = \delta_i^j +q \hat{R}^{jk}_{il}x^l \pa_k.
\label{ap9}
\ee
The commutation relations between $x^i$ and $\xi^j$ are
\be
x^i \xi^j = q \hat{R}^{ij}_{kl}\xi^k x^l.
\label{ap10}
\ee
We need also the commutation relations between $\pa_i$ and $\xi^j$:
\be
\pa_i \xi^j = q^{-1} \hat{R}^{-1\ jk}_{\ \ \ \ il}\xi^l \pa_k.
\label{ap11}
\ee
The algebra of the derivatives is
\be
 P^{- \ ij}_{\ \ \ kl} \pa_j \pa_i= 0 \ .
\label{ap12}
\ee
The element
\be
\de=\frac{1}{1+q^{N-2}} g^{ij}\pa_j \pa_i
\label{ap13}
\ee
is central in
the algebra of the derivatives, $\de\pa_i=\pa_i\de$.

3. The compact form of $SO_q(N)$ is defined for a real $q$.
In this case we
have $\overline{\hat{R}}=\hat{R}$. The conjugation
of the coordinates has a form
\be
\overline{x^i}=g_{ji}x^j.
\label{ap14}
\ee
It defines the euclidean q-space. The length $L$ is real under
this conjugation, $\overline{L}=L$.

\sectione{Conjugate Derivatives}
\vskip 0.3cm

In this section
we find the action of the conjugate derivatives and express
the conjugate derivatives in terms of the
derivatives themselves. Also, we
construct a real Laplacian.

According to \cite{WZCOV} the covariant and consistent
derivatives are defined by the expression (\ref{ap9})
involving $\hat{R}$-matrix.
One can define another set of consistent
and covariant derivatives using $\hat{R}^{-1}$ instead. First of
all we show that in the q-orthogonal case this gives the
conjugate derivatives.

{\bf Lemma}.
\be
\hat{\partial_k}x^v =
\delta_k^v+q^{-1}\hat{R}^{-1\ vi}_{\ \ \ \ kj}
x^j\hat{\partial_i},
\label{2p1}
\ee
where
\be
\hat{\partial_i}=-q^N g_{ik}g^{tk}\overline{\partial^t}.
\label{2p2}
\ee

{\it Proof}. To write relations conjugated to (\ref{ap9}) in the
form (\ref{2p1}) one finds first a tensor
$\Phi^{ks}_{nv}$, inverse to $\hat{R}^{va}_{sb}$ in indices
$(v,s)$. Put
\be
\Phi^{ks}_{nv}=g_{nl}\hat{R}^{kl}_{uv}g^{us}.
\label{2p3}
\ee
Using relations (\ref{ap3}) one finds
\be
\Phi^{ks}_{nv}\hat{R}^{va}_{sb}=
\hat{R}^{ks}_{nv}
\Phi^{va}_{sb}=\delta^k_b \delta^a_n .
\label{2p4}
\ee
Now one proves (\ref{2p1}) by a straightforward calculation
using (\ref{ap14}) and (\ref{2p2}).

Comparing (\ref{2p1}) with (\ref{ap9})
one sees that the derivatives
$\partial_i$ and $ \hat{\partial_i}$ act
in the same way on the linear
functions of $x^j$ but their actions on
higher order polynomials do not
coincide. Therefore the conjugate derivatives cannot be expressed
linearly in terms of the derivatives themselves. It turns out that
$\hat{\partial}_i$ can be expressed nonlinearly in $\partial_j$.
To write this expression we need the scalar
operators $L,\Delta$ and
$E=x^i\partial_i$. The commutation relations of these operators
with the coordinates and the derivatives are
\bea
Lx^k&=& x^k L, \\
[0.5ex]
\pa_k L &=&q^2 L\pa_k +
q^{2-N} x_k, \\
[0.5ex]
\de x^k&=&q^2 x^k \de + q^{2-N}\pa^k ,   \\
[0.5ex]
\pa_k\Delta &=& \Delta \pa_k ,\\
[0.5ex]
E x^k&=&q^2 x^k E + x^k - q\lambda L \pa^k,\\
[0.5ex]
\pa_k E &=& q^2 E\pa_k +\pa_k -q\l x_k\Delta. \\
\label{2p5}
\eea
Finally we will use the operator $\L$, introduced in \cite{OG}:
\be
\L = 1+ q\l E +q^N \l^2 L\de \ .
\label{2p8}
\ee
It obeys homogeneous relations with both the coordinates and the
derivatives,
\be
\L x^k = q^2 x^k \L \ ,\ \L \pa_k = q^{-2}\pa_k \L \ \ .
\label{2p9}
\ee
Now we are ready to formulate the main result of this section.

{\bf Theorem}.
\be
\hat{\pa_k}=q^{N-2}\L^{-1}[\de , x_k].
\label{2p10}
\ee

{\it Proof}. Denote
$T_k=q^{N-2}[\de ,x_k]$. Using (\ref{2p5}) we
can write
\be
T_k = \pa_k + q^{N-1}\l x_k \de \ .
\label{2p11}
\ee
Compute
$$
T_ix^j-q\hat{R}^{-1\ jk}_{\ \ \ \ il}x^l T_k =
$$
\be
\dd^j_i+ q\hat{R}^{jk}_{il}x^l\pa_k
+q^{N+1}\l x_i (x^j\de + q^{-N}\pa^j)-
q\hat{R}^{-1\ jk}_{\ \ \ \ il}x^l\pa_k -
q^N\l \hat{R}^{-1\ jk}_{\ \ \ \ il}x^l x_k \de \ .
\label{2p12}
\ee
For the terms with $xx\de$ we have
$$
qx_i x^j-
\hat{R}^{-1\ jk}_{\ \ \ \ il}x^l x_k \de =
$$
$$
g_{ia}(
qx^a x^j-g^{au}
\hat{R}^{-1\ jk}_{\ \ \ \ ul}g_{ks}x^l x^s ) =
$$
\be
g_{ia}(q\ID - \hat{R})^{aj}_{ls}x^l x^s =
\label{2p13}
\ee
$$
g_{ia}(q-q^{1-N})P^{o\ aj}_{\ \ ls} x^l x^s = \dd^j_i\l L \ .
$$
In the
second equality we used (\ref{ap3}). In the third equality
we used
(\ref{ap1}), (\ref{ap5}), and the completness of the set of the
projectors $P^+,P^-,P^o$,
\be
\ID = P^+ +P^- +P^o \ .
\label{2p14}
\ee
In the fourth - equations
(\ref{ap2}) and (\ref{ap6}) were used. For the terms with
$x\pa$ we have
$$
(q\hat{R}^{jk}_{il}-
q\hat{R}^{-1\ jk}_{\ \ \ \ il}
+q\l g_{il}g^{jk})x^l\pa_k =
$$
\be
((q^2-1)P^+ + (q^2 -1)P^-
+(q^{2-N}-q^N +q\l \nu^{-1})P^o)^{jk}_{il}
x^l \pa_k =
\label{2p15}
\ee
$$
q\l\dd^j_i E \ ,
$$
where $\nu$ is given by (\ref{ap2}).
Here in the first equality we used (\ref{ap1}) and (\ref{ap2}),
and in the second - (\ref{2p14}). Collecting
all terms together we obtain
\be
T_ix^j=\dd^j_i \L +q\hat{R}^{-1\ jk}_{\ \ \ il}x^l T_k \ .
\label{2p16}
\ee
Now multiplying by $\L^{-1}$ from the left, using (\ref{2p9})
and comparing with (\ref{2p1}) we conclude that the lhs and rhs
of (\ref{2p10}) have the same commutation relations with $x^i$,
which completes the proof.

We note that although the conjugation rule (\ref{2p10}) is
nonlinear, on conjugating twice all nonlinearities
disappear and $\overline{\overline{\pa_i}}=\pa_i$.
The map inverse to (\ref{2p10}) is
\be
\pa_k = (\overline{\L})^{-1}[\overline{\de},x_k]\equiv
\L(\hat{\pa}_k-q^{N-3}\l x_k \L^{-1} \de ).
\ee
To complete the
treatment we find the reality properties of the operators
$E,\de , \L $. By a
somewhat lengthy but straightforward calculation
one finds:
\be
\overline{\de}=q^{-N-2}\L^{-1}\de \ ,
\label{2p17}
\ee
\be
\overline{E}=-q^{-N}\L^{-1}((q^N+q\l)E+
\frac{(q^N-1)(q^{1-N}+q^{-1})}{\l} +
q^{N+1}\l (1+q^{N-2})L\de ) \ .
\label{2p18}
\ee
Therefore, using (\ref{deltal}), we obtain
\be
\overline{\L} = q^{-2N}\L^{-1} \ .
\label{2p20}
\ee
Equation (\ref{2p17})
shows that the Laplace operator $\de$ built out
of the derivatives $\pa_i$ only is not real. However, for
\be
\de_R = \L^{-1/2}\de
\label{2p21}
\ee
we have
\be
\overline{\de_R}=\de_R \ .
\label{2p22}
\ee
Therefore, $\de_R$ is a good candidate for a real Laplacian.

\sectione{Conjugate Differentials}
\vskip 0.3cm

In this section
we express the conjugate differentials in terms of
the differentials themselves, and find the reality property of
the exterior derivative.

Conjugating the relation (\ref{ap11}) and defining
$\hat{\xi}^i$ by
\be
\overline{\xi^j} = g_{ij}\hat{\xi}^i \ ,
\label{3p1}
\ee
we find using (\ref{ap1}),
\be
\hat{\xi}^i x^j =q \hat{R}^{ij}_{kl}x^k \hat{\xi}^l \ .
\label{3p2}
\ee
To find the relation between $\hat{\pa}_k$ and $\pa_l$ we used
scalar operators
obtained by contraction of indices of $x^i,\pa_j$.
Now we need two more scalar operators, the exterior derivative
$d$ and the operator
\be
W=\xi^i x_i \ .
\label{3p3}
\ee
The commutation
relations of $L,\de , E$ with $\xi^i$ are simple:
\bea
 L \xi^i&=&q^2 \xi^i L\ , \\
[0.5ex]
 \de \xi^i&=&q^{-2} \xi^i \de \ ,\\
[0.5ex]
 E \xi^i& =& \xi^i E \ .
\label{3p4}
\eea
The new operators $d$ and $W$ have the following commutation
relations with $x^i$, $\pa_i$, and $\xi^i $:
\bea
d x^i&=&\xi^i +x^i d\ ,\\
[0.5ex]
d\partial_i&=&q^2  \partial_i d- q^{N-1}\l \xi_i \de \ ,\\
[0.5ex]
d\xi^i&=&-\xi^i d \ ,\\
[0.5ex]
 W x^i&=&x^i W - q^{N-1}\l \xi^i L \ ,\\
[0.5ex]
 \partial_j W&=&W \partial_j + q^{N-2}\xi_j - q^{-1}\l x_j d
 + q^{N-1} \l \xi_j E \ ,\\
[0.5ex]
 W \xi^j&=&-q^2 \xi^j W \ .
\label{3p5}
\eea
As for relations (\ref{2p5}) we leave
a check of these
relations to the reader. This check can be reduced to
manipulations with
the symmetry properties and projector decomposition of
the $\hat{R}$-matrix.

We introduce also a quantity
\be
U=W+q^{N-3}\l L d \ ,
\label{3p7}
\ee
which commutes with all coordinates,
\be
Ux^i = x^i U \ ,
\label{3p8}
\ee
while with the derivatives and differentials it obeys
\bea
\pa_j U & = & U\pa_j+q^{N-2}\xi_j\L \ ,\\
[0.5ex]
\xi^j U & =& -q^{-2}U\xi^j \ .
\eea
The commutation relations of $\xi^i$ with $x^j$ or $\pa_k$ are
homogeneous. Hence rescaling $\xi^i$ by a
numerical factor does not change any
of them. This
shows that the commutation relations with $x^j$ imply
the expression
of $\hat{\xi}^l$ in terms of $\xi^i$ only up to a
factor. Demanding
the square of conjugation to be unity, one fixes
the absolute value of this factor.

{\bf Theorem}.
\be
\hat{\xi}^i = \sigma q^N \L (\xi^i + q^{-1}\l x^i d
-q^{1-N}\l U \hat{\pa}^i ) \ ,
\label{3p9}
\ee
where $\sigma$ is a pure phase, $\sigma = e^{i\phi}$.

{\it Proof}. Again,
once the rhs of (\ref{3p9}) is written, one can check
(using the
projector
decomposition and the symmetries of the $\hat{R}$-matrix)
that it has the same commutation relations with $x^i$
as $\hat{\xi}^j$ do.

To prove that $\sigma$ is a pure phase, we find the square
of the conjugation. To this end we need the expressions
for $\overline{d}$ and $\overline{W}$. A straightforward
calculation shows that
\be
\overline{d}=-\sigma q^N (\L d - q\l U \de),
\label{3p10}
\ee
and
\be
\overline{W}=\sigma q^N (q^{2-N} U + q^{N-5}\l\L L d -
q^N \l^2 U L \de ) \ .
\label{3p11}
\ee
It then follows that
\be
\overline{U}=\sigma U \ .
\label{3p12}
\ee
Conjugating
(\ref{3p12}) we find that $\sigma$ is a pure phase as
stated. One
more check shows that the square of the conjugation is
unity on $\xi^i$ as well. This finishes the proof.

The mapping inverse to (\ref{3p9}) is
\be
\xi^i=\overline{\sigma}q^N\overline{\L}(\hat{\xi}^i +\l q
x^i \overline{d} +q^3\l \overline{U}\pa^i) .
\ee

As in the discussion of the reality properties of the Laplacian,
one may build a combination of operators which reduces to $d$ in
the classical limit and has a linear conjugation law. One
choice is
\be
d_{\scriptstyle 0} = \L^{1/2}d+(1-q)\L^{-1/2}U\de \ .
\ee
Then
\be
\overline{d_{\scriptstyle 0}}=-\sigma d_{\scriptstyle 0} \ .
\label{dconj}
\ee
However this choice destroys the fundamental nilpotency property
of the exterior derivative. Another possibility is simply
to take
\be
d_{\scriptstyle 1}=\frac{1}{2}(d-\overline{\sigma}\overline{d})=
\frac{1}{2}((1+q^N\L)d-q\l U\de)\ .
\ee
Then
\be
\overline{d_{\scriptstyle 1}}=
-\sigma d_{\scriptstyle 1}\ ,
\ee
as in (\ref{dconj}). Moreover, one finds
\be
d\overline{d}+\overline{d}d=0\ .
\ee
Therefore
\be
d_{\scriptstyle 1}^2=0\ .
\ee

We note
that rescaling $\xi^i$ by a phase one can eliminate the
factor $\sigma$ in the above formulas.

To conclude, we
stress once more that the mappings (\ref{2p10}) and
(\ref{3p9}) are covariant under the quantum group $SO_q(N)$.

{\it Remark}. In the SL$_q(n, I\!\! R)$ case
the action of the conjugate derivatives is given by
the $\hat{R}$-matrix itself and therefore the conjugation rules
for the derivatives are linear \cite{WZCOV}, \cite{JWTALK}.
However still there is another set of covariant and consistent
derivatives defined with the help of $\hat{R}^{-1}$,
\be
\pa'_i  x^j =
\delta^j_i +q^{-1}\hat{R}^{-1\ jk}_{\ \ \ \ il}x^l\pa'_k \ ,
\label{dprime}
\ee
and one may ask
how they are related to the original ones. Using
the relation
\be
\hat{R}=\hat{R}^{-1}+\l \ ,
\ee
valid in the SL case, we may rewrite the action of the original
derivatives in the form
\be
\pa_i x^j=\delta^j_i+q\hat{R}^{-1\ jk}_{\ \ \ \ il}x^l\pa_k +
q\l \delta^j_i x^l\pa_l =\mu_1\delta^j_i
+q\hat{R}^{-1\ jk}_{\ \ \ \ il}x^l\pa_k \ ,
\label{slmap}
\ee
where $\mu_1 = 1+q\l E$. The
operator $\mu_1$ is multiplicative \cite{OG}:
\be
\mu_1 x^i =
q^2 x^i \mu_1 \ , \ \ \mu_1 \pa_i = q^{-2}\pa_i \mu_1 \ .
\ee
Multiplying
(\ref{slmap}) by $\mu_1^{-1}$ from the left we see that
the operators $\mu_1^{-1}\pa_i$ satisfy
the same commutation relations with
the coordinates as $\pa'_i$. Therefore we may set
\be
\pa'_i = \mu_1^{-1}\pa_i \ .
\ee
This is the needed relation between the original and primed
derivatives.

The primed differentials, defined by
\be
x^i \xi'^j = q^{-1}\hat{R}^{-1\ ij}_{\ \ \ \ kl}\xi'^k x^l \ ,
\ee
can be expressed in terms of the original differentials as well.
One checks that the quantities $\mu_1(\xi^i + q^{-1}\l x^i d)$
satisfy the same commutation relations with the coordinates as
$\xi'^i$. As in the q-orthogonal case this gives
the relation between the primed and original differentials up
to an overall numerical factor $c$,
\be
\xi'^i=c\mu_1(\xi^i + q^{-1}\l x^i d) \ .
\ee
One now checks that $\mu_1$
commutes with $\xi^i$. Therefore we also find the
relation
\be
d'\equiv \xi'^i\pa'_i=
c\mu_1(\xi^i+q^{-1}\l x^id)\mu_1^{-1}\pa_i=
c(d+q\l Ed)=c\mu_1d
\ee
between the primed and original exterior derivatives.
\startapp

\sectione{Useful Formulas}
\vskip 0.3cm

Here we collect various identities and commutation relations
needed for
checks and proofs of the statements in Sections 2 and 3.

The action
of the scalar operators on the coordinates, derivatives and
differentials
was given in the text. Here are some useful commutators
between
the scalar operators themselves.
 \bea
 E L&=&q^2 L E + (q^{2-N}+1)L\ ,\\
[0.5ex]
\de E&=&q^2 E \de +(q^{2-N}+1)\de\ ,\\[0.5ex]
\de L&=&q^4L\de +
q^{4-N}E +\fracds{q^{2-N}-q^{2-2N}}{q^{-1}\l }\ ,\\
[0.5ex]
 d L&=&q^{-N+2}W + L d\ ,\\
[0.5ex]
 d W&=&- W d\ ,\\
[0.5ex]
d E&=& q^2E d +d- q \l W \de\ ,\\
[0.5ex]
 \de W&=&W \de + q^{-N} d\ ,\\
[0.5ex]
 \de U&=&U \de + \L d\ ,\\
[0.5ex]
\de d&=&q^{-2}d\de\ ,\\
[0.5ex]
dU&=&-q^{-2}Ud\ .
\label{deltal}
\eea
Also, the following summation rules
\bea
\pa^i x_i&=&\fracds{(q^N-1)(q^{1-N}+q^{-1})}{\l } +q^N E\ ,\\
[0.5ex]
\pa^i\xi_i&=&q^{-N}d\ ,\\
[0.5ex]
x^i\xi_i&=&q^{2-N}W\ ,\\
[0.5ex]
\hat{\xi^i}\hat{\pa_i}&=&-\overline{d}\ ,\\
[0.5ex]
\hat{\pa^i}\hat{\pa_i}&=&(1+q^{N-2})q^{N-2}\L^{-1}\de
\eea
were used.

{\it Remark}. The operators
\be
e=q^{N-1} L\ , \ \ h=q^N (E+\frac{1-q^{-N}}{q\l})
\ \ {\rm and}\ \ f=q^{N-1} \de
\ee
satisfy the relations of the q-deformed
$sl(2)$-algebra,
\bea
q^{-1}he-q eh&=&(q+q^{-1})e \ \ ,\\
q^{-1}fh-qhf&=&(q+q^{-1})f \ \ ,\\
q^{-2}fe-q^2ef&=&h \ .
\eea
Note that the operator entering $\L$,
\be
q^N\L = 1+\l \tilde{h}\ \ ,\ \ \tilde{h}=qh+q^2\l ef
\ee
has an algebraic meaning as well. We have
\bea
q^{-2}\tilde{h}e-q^2 e\tilde{h}&=&(q+q^{-1})e \ \ ,\\
q^{-2}f\tilde{h}-q^2 \tilde{h}f&=&(q+q^{-1})f \ \ ,\\
q^{-1}fe-q ef&=&\tilde{h} \ ,
\eea
which is another form of the
$sl_q(2)$-algebra.

\vskip 0.2cm

\vskip 0.75 cm
{\bf Acknowledgements}. It is a
pleasure for us to thank J. Bobra,
H. Ewen, V.
Jain and W. Schmidke
 for valuable discussions. We are
 especially grateful to J. Wess,
discussions with
whom brought up questions treated in the paper.
\vskip 0.3cm
\vfill

\end{document}